\documentclass[times,preprint,authoryear]{elsarticle}

\usepackage{jasr}
\usepackage{framed,multirow}

\usepackage{amssymb}
\usepackage{latexsym}

\usepackage[switch]{lineno}

\usepackage{url}
\usepackage{xcolor}
\definecolor{newcolor}{rgb}{.8,.349,.1}

\usepackage[citebordercolor=white]{hyperref}

\journal{Advances in Space Research}

\begin{document}

\verso{Kumiko K. Nobukawa \textit{etal}}

\begin{frontmatter}

\title{Confirmation of dust scattering echo around MAXI\,J1421$-$613  \\ by Swift observation}%

\author[1]{Kumiko K. \snm{Nobukawa}\corref{cor1}}
\cortext[cor1]{Corresponding author}
\ead{kumiko@phys.kindai.ac.jp}
\author[2]{Masayoshi \snm{Nobukawa}}
\author[3]{Shigeo \snm{Yamauchi}}

\address[1]{Faculty of Science and Engineering, Kindai University, 3-4-1 Kowakae, Higashi-Osaka, 577-8502, Japan}
\address[2]{Faculty of Education, Nara University of Education, Takabatake-cho, Nara, Nara, 630-8528, Japan}
\address[3]{Faculty of Science, Nara Women's University, Kitauoyanishi-machi, Nara, Nara, 630-8506, Japan}

\received{}
\finalform{}
\accepted{}
\availableonline{}
\communicated{}

\begin{abstract}
MAXI\,J1421$-$613 is an X-ray burster discovered by Monitor of All-sky X-ray Image (MAXI) on 9 January 2014 and is considered to be a low-mass X-ray binary.   A previous study analyzing follow-up observation data obtained by Suzaku on 31 January to 3 February 2014 reported that an annular emission of $\sim3'$--$9'$ radius was found around the transient source.  The most plausible origin of the annular emission is a dust scattering echo by the outburst of MAXI\,J1421$-$613. In this paper, we confirm the annular emission by analyzing the data of the Swift follow-up observation which was conducted by the photon counting mode on 18 January 2014. In a radial profile, we found an annular emission at $\sim2'.5$--$4'.5$. Its spectrum was well explained by an absorbed power law, and the photon index was higher than that of MAXI\,J1421$-$613 itself by $\Delta \Gamma \sim 2$. The flux and radius of the annular emission observed by Swift are explained by dust scattering of the same outburst as is responsible for the annular emission observed by Suzaku.  Assuming that the dust layer causing the annular emission found by Swift is located at the same position as the CO cloud in front of MAXI\,J1421$-$613, the distance to the transient source was estimated to be $\sim3$~kpc, which is consistent with the value estimated by the previous study of Suzaku. 
\end{abstract}


\end{frontmatter}


\section{Introduction}
\label{sec1}
MAXI\,J1421$-$613 is a neutron-star low-mass X-ray binary that was discovered by the MAXI Nova Alert System \citep{Negoro10} on 9 January 2014 \citep{Morooka14}.   \citet{Serino15} analyzed the data of the MAXI Gas Slit Camera  (GSC)  and a series of Swift X-Ray Telescope (XRT) follow-up observations. According to the authors, MAXI/GSC recorded that the source started brightening around 7 January 2014 and exhibited a flux peak with $\sim10^{-9}$~erg~cm$^{-2}$~s$^{-1}$ around 9--10 January 2014. During the outburst, the Swift/XRT spectra can be explained by thermal Comptonization of multi-color disk blackbody emission with a typical photon index for low-mass X-ray binaries, $\Gamma \sim2$.  The authors also analyzed the Suzaku data which were taken on 31 January to 3 February 2014 and obtained the $3\sigma$ upper limit flux of $1.2\times10^{-13}$~erg~cm$^{-2}$~s$^{-1}$.

\citet{Nobukawa20} discovered an annular emission of $\sim3'$--$9'$ radius around the center of MAXI\,J1421$-$613 in the Suzaku data. 
The authors also analyzed the swift data taken in almost the same period as Suzaku and found a hint of a peak of $\sim6'$ in its radial profile. 
The spectrum of the annular emission is well explained by an absorbed power law with a photon index of $\Gamma\sim4.2$, which is too soft for the non-thermal emission of shell-type supernova remnants.  The observed $\Gamma$ of the annular emission is higher than that of MAXI\,J1421$-$613 ($\Gamma\sim 2.1$; Serino et al., 2015) by $\Delta\Gamma \sim 2$. Also, a hint of the radius expansion from $\sim5'.4$ to $\sim5'.9$ and a hint of flux decrease from $2.6\times10^{-12}$~erg~cm$^{-2}$~s$^{-1}$ to $2.2\times10^{-12}$~erg~cm$^{-2}$~s$^{-1}$ were found in the Suzaku three-day long observation. These results support the dust scattering echo scenario as the origin of the annular emission. 
The authors found a spatial correlation between the CO maps from \citet{Dame01} and the distribution of the hydrogen column density of the annular emission.  This means that MAXI\,J1421$-$613 is behind a CO cloud. Putting an assumption that the dust layer is located at the same position as the CO cloud, the authors concluded that the distance to MAXI\,J1421$-$613 is $\sim3$~kpc.

In this paper, we examined the Swift data not analyzed in \citet{Nobukawa20} and confirmed the annular emission at a different radius from the Suzaku result. 
We found that all data of the annular emission obtained by Suzaku and Swift can be explained consistently as a dust scattering echo of the outburst of MAXI\,J1421$-$613 which peaks on 9--10 January 2014.

\section{Observation Data}
Within the same day as the MAXI alert on 9 January 2014, Swift started follow-up observations of MAXI\,J1421$-$613 with the photon counting (PC) mode (ObsID$=$00033092001, 00033093001). 
 After them, Swift conducted a series of observations
with eleven segments from 11 January to 4 February 2014 (ObsID$=$00033098001--0033098011). All segments except for segments 9 and 10 were obtained by windowed timing (WT) mode. 
During segment 5, an X-ray burst occurred and the Burst Alert Telescope (BAT) was triggered \citep{Baumgartner14}, so a follow-up observation with the PC mode was started (ObsID=00584155000). \citet{Serino15} referred to this observation as segment 5$+$. 
We summarized the Swift observations in table~\ref{tb:Obs}.  

The observation periods of  ObsID$=$00033098009 and 00033098010 overlapped with that of Suzaku. \citet{Nobukawa20} analyzed these data sets and found a hint of a peak at $\sim6'$ radius in the radial profile, which is consistent with the Suzaku result. 
In this paper, we analyzed the remaining PC mode observations of ObsID$=$00033092001, 00033093001, and 00584155000.  

We also analyzed Swift/XRT data of another source obtained by the PC mode as a reference for a radial profile. 
We chose the low-mass X-ray binary 4U\,1323$-$619 (ObsID=00087249047) since it is bright but not enough to cause pileup, and persistent \citep{Warwick81}. In fact, according to the Swift/BAT transient monitor (https:\slash\slash swift.gsfc.nasa.gov\slash results\slash transients\slash index.html),  the light curve of 4U\,1323$-$619 showed the time variability of only $\sim10$\% from the observation date of the analyzing data to 50 days before. The observation log of 4U\,1323$-$619 is shown in table~\ref{tb:Obs}. 

We used the analysis software package HEAsoft 6.28.  Errors in this paper are quoted at the 90\% confidence levels while the error bars in the figures show the 1$\sigma$ errors.

\begin{table*}
\centering
\caption{Swift/XRT observation logs. The observations with the asterisk marks ($^*$) were analyzed in this paper. The dagger marks ($^\dag$) observations were analyzed in \citet{Nobukawa20}. }\label{tb:Obs}
\begin{tabular}{llcccccc} \hline
Object	& ObsID   	& \multicolumn{2}{c}{Pointing direction}                                                        & \multicolumn{2}{c}{Observation time (UT)}             & XRT Exposure (s)  \\
	 	&		&   $\alpha_{\rm J2000.0} (^{\circ})$ & $\delta_{\rm J2000.0} (^{\circ})$  	& Start  				& End 				& in PC mode  \\	
\hline
MAXI\,J1421$-$613 	& 00033092001$^*$	& $215^{\circ}.76$	& $-61^{\circ}.64$			& 2014-01-09 19:49:06	& 2014-01-09 22:14:32	& 461	\\
				& 00033093001$^*$	& $215^{\circ}.05$	& $-61^{\circ}.64$			& 2014-01-09 19:55:09	& 2014-01-09 22:11:59	& 461	\\
				& 00033098001	& $215^{\circ}.38$	& $-61^{\circ}.61$			& 2014-01-11 05:34:58	& 2014-01-11 08:07:13	& 0	\\
				& 00033098002	& $215^{\circ}.38$	& $-61^{\circ}.61$			& 2014-01-13 02:28:59	& 2014-01-13 04:47:42	& 0	\\
				& 00033098003	& $215^{\circ}.36$	& $-61^{\circ}.58$			& 2014-01-14 19:49:58	& 2014-01-14 20:43:37	& 0	\\
				& 00033098004	& $215^{\circ}.42$	& $-61^{\circ}.60$			& 2014-01-16 21:32:11	& 2014-01-16 23:48:29	& 0	\\
				& 00033098005	& $215^{\circ}.39$ 	& $-61^{\circ}.61$			& 2014-01-18 08:30:59	& 2014-01-18 09:23:48	& 0	\\
				& 00584155000$^*$ 	& $215^{\circ}.38$  	& $-61^{\circ}.61$   			& 2014-01-18 08:41:09  	& 2014-01-18 12:20:12 	& 6068      \\ 
				& 00033098006	& $215^{\circ}.40$	& $-61^{\circ}.61$			& 2014-01-20 08:44:59	& 2014-01-20 11:03:47	& 0	\\
				& 00033098007	& $215^{\circ}.41$	& $-61^{\circ}.59$			& 2014-01-23 08:51:59	& 2014-01-23 09:43:59	& 0	\\
				& 00033098008	& $215^{\circ}.39$	& $-61^{\circ}.60$			& 2014-01-28 23:32:59	& 2014-01-29 00:24:58	& 0	\\
				& 00033098009$^\dag$	& $215^{\circ}.48$	& $-61^{\circ}.59$			& 2014-01-30 20:33:41	& 2014-01-30 22:59:21	& 552	\\
				& 00033098010$^\dag$	& $215^{\circ}.41$	& $-61^{\circ}.57$			& 2014-02-01 07:29:59	& 2014-02-01 10:09:32	& 1975	\\
				& 00033098011		& $215^{\circ}.44$	& $-61^{\circ}.56$			& 2014-02-03 17:26:58	& 2014-02-04 00:35:29	& 0	\\
\hline
4U\,1323$-$619 	 & 00087249047$^*$ 	&  $201^{\circ}.72$   	& $-62^{\circ}.13$			& 2017-10-22 01:18:30    	& 2017-10-23 23:59:53	& 7240        \\
\hline
\end{tabular}
\end{table*}

\section{Analysis and Results}
\subsection{X-ray images and radial profiles}
Figure~1(a) is an X-ray image of MAXI\,J1421$-$613 in the 1--5~keV band obtained by Swift/XRT on the same day as the MAXI alert on 9 January 2014. Two observations of ObsID$=$00033092001 and 00033093001 were merged. Figure~1(b) is the same as (a), but the observation of ObsID$=$00584155000 conducted on 18 January 2014. In the latter, besides MAXI\,J1421$-$613, which is the bright source at the center, faint and diffuse X-ray emission is seen between the two circles which have radii of $2'.5$ and $4'.5$ centered on MAXI\,J1421$-$613. This diffuse emission may be a dust scattering echo of MAXI\,J1421$-$613. For comparison, an X-ray image of 4U\,1323$-$619 was also produced in the 1--5~keV band (figure 1c).  The circles of radii $2'.5$ and $4'.5$ from 4U\,1323$-$619 are overlaid on the image. There is little emission outside the circle of radius of $2'.5$.

\begin{figure}[t]
\centering
\includegraphics[scale=0.215]{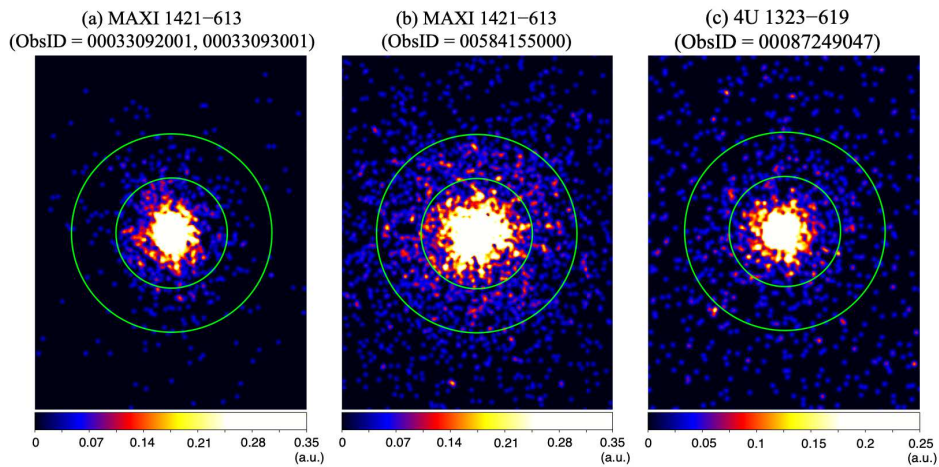}
\caption{(a) X-ray image of MAXI\,J1421$-$613 in the 1--5~keV band obtained by Swift/XRT on the same day as the MAXI alert on 9 January 2014. Two observations of ObsID$=$00033092001 and 00033093001 were merged. Green circles represent radii $2'.5$ and $4'.5$ from MAXI\,J1421$-$613. 
(b) Same as (a) but of the observation of ObsID$=$00584155000 on 18 January 2014. A spectrum of the annular emission was extracted from the region bounded by the two circles with radii of $2'.5$ and $4'.5$ from MAXI\,J1421$-$613. (c) X-ray image of 4U1323$-$619 in the 1--5~keV band obtained by Swift/XRT. Green circles represent radii $2'.5$ and $4'.5$ from 4U\,1323$-$619.}
\label{fig:pendulum}
\end{figure}

We made a radial profile of MAXI\,J1421$-$613  in the 1--5~keV band for the former observations (ObsID$=$00033092001 and 00033093001) as shown in figure 2(a). The effect of pileup is severe in these observations, and thus count rate is decreasing inside the radius of $\sim0.1'$. 
Even if a ring echo were to occur during the observations, it is difficult to constrain the angular size because of the pileup.

Figure 2(b) presents radial profiles of MAXI\,J1421$-$613 for the observation on 18 January 2014 (ObsID$=$00584155000)  and 4U\,1323$-$619 for reference. 
They seem to be free from the pileup.
Since the PSF of Swift/XRT is known to be represented by a King model  \citep{Moretti05}, we fitted the two radial profiles simultaneously with a King model representing the PSF in addition to a constant component representing the background as follows:
\begin{eqnarray}
I (r) = A\times\left(1+\left(\frac{r}{r_c}\right)^2\right)^{-\beta}+C. 
\end{eqnarray}
4U\,1323$-$619 is offset from the pointing center by $1'.98$, while MAXI\,J1421$-$613  is offset from the pointing center by $0'.71$.  
The differences of $r_c$ and $\beta$ values between offsets of $\sim0'.7$ and $\sim2'$ in the PSF file of the Swift XRT CALDB \citep{Moretti05} are within the statistical errors of our fitting. Also, we found the constant parameters $C$ of the two profiles were consistent within the statistical errors. Therefore the values of $r_c$, $\beta$, and $C$ were free but linked between the two profiles. 
The normalization $A$ was free independently for the two profiles. 
 The model basically reproduces both profiles, but an excess at $r\sim2'.5$--$4'.5$ exists only in the radial profile of MAXI\,J1421$-$613. This feature should be caused by the annular emission seen in figure 1(b).  Therefore, we added a Lorentzian to the fitting model of the radial profile of MAXI\,J1421$-$613  and simultaneously fitted the two radial profiles again.  Here, the parameters of the King model and the constant component were fixed to the values obtained in the previous fitting whereas the parameters of the Lorentzian were free. This model reproduced the radial profile of MAXI\,J1421$-$613 well (figure 2c). The centroid of the Lorentzian was obtained to be $3'.5\pm0'.2$.

\begin{figure*}
\centering
\includegraphics[scale=0.077]{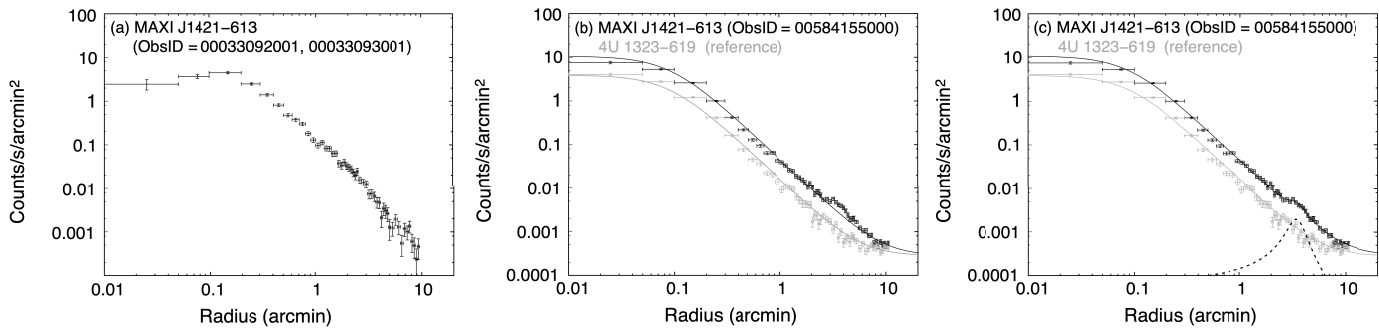}
\caption{(a) Radial profile of MAXI\,J1421$-$613 of ObsID = 00033092001 and  00033093001 in the 1--5~keV band obtained by Swift/XRT. The count rate is decreasing inside the radius of $\sim0.1'$ due to pileup. 
(b) Radial profiles of MAXI\,J1421$-$613 of ObsID = 00584155000 (black) and 4U\,1323$-$619 (gray) in the 1--5~keV band. The fitted model consists of a King model and a constant component. The solid black and gray lines indicate the best-fit models for MAXI\,J1421$-$613 and 4U\,1323$-$619, respectively. 
(c) Same as (b), but the fitted model for MAXI\,J1421$-$613 contains a Lorentzian (dashed line) as well as the King model and the constant component.  }
\label{fig:pendulum}
\end{figure*}

\subsection{Spectral analysis}
Since the annular emission was clearly seen in the observation performed on 18 January 2014 (ObsID$=$00584155000), we performed spectral analysis for this observation.  First, we analyzed a spectrum of MAXI\,J1421$-$613 to measure its flux. The spectrum was extracted from a circle of radius $2'$ centered on the point source. As background, a spectrum was extracted from an annular region of radii $6'$--$10'$ centered on MAXI\,J1421$-$613 (outside the view of figure1b). The background-subtracted spectrum of MAXI\,J1421$-$613 was fitted with an absorbed thermal Comptonization of multi-color disk blackbody emission ({\tt wabs$\times$nthComp} in XSPEC), which is the same model as the previous study \citep{Serino15}. All the parameters but flux were fixed to those in  \citet{Serino15}; $N_{\rm H} = 4.8 \times10^{22}$~cm$^{-2}$, $kT_{\rm in}=0.37$~keV, $\Gamma=2.1$, and $kT_{\rm e}=20$~keV. The flux of MAXI\,J1421$-$613 was obtained to the  $1.3\times10^{-10}$~erg~s$^{-1}$~cm$^{-2}$ in the 0.5--10~keV band.  Reduced chi-squared was $\chi^2/$d.o.f.$= 326.16/211$.
We modified the absorption model {\tt wabs} to {\tt phabs} in XSPEC, keeping the parameters of {\tt nthComp}  fixed to those of \citet{Serino15},  and fitting the spectrum with free $N_{\rm H}$ value and flux. 
Reduced chi-squared was $\chi^2/$d.o.f.$= 331.72/209$, and the best-fit value of $N_{\rm H}$ was $(4.6\pm0.1)\times10^{22}$~cm$^{-2}$. The flux of MAXI\,J1421$-$613 was measured to be $1.3\times10^{-10}$~erg~s$^{-1}$~cm$^{-2}$ in the 0.5--10~keV band. 

Then, we analyzed the annular emission. A spectrum was obtained from the region bounded by the two circles in figure 1(b): the annular region of radii $2.5'$--$4.5'$ centered on MAXI\,J1421$-$613. The background spectrum was the same in the spectral analysis of MAXI\,J1421$-$613.  The two spectra are shown in figure~3. 
We fitted the background-subtracted spectrum in the 1.5--6.0~keV band.  As seen in figure~2(c), the PSF of MAXI\,J1421$-$613 is still dominant in the annular region of radii $2.5'$--$4.5'$.   
To take into account the PSF as well as the annular emission, we generated two ARF files by {\tt xrtmkarf}: for MAXI\,J1421$-$613 (a point source) and the annular emission (an extended source). Applying the two different ARFs to the two models follows the XSPEC manual (https:\slash\slash heasarc.gsfc.nasa.gov\slash xanadu\slash xspec\slash manual\slash node40.html). 
The applied fitting model was an absorbed thermal Comptonization of multi-color disk blackbody emission ({\tt phabs$\times$nthComp} in XSPEC) plus an absorbed power law ({\tt phabs$\times$powerlaw}). The former represents MAXI\,J1421$-$613, and the latter represents the annular emission and is the same model as in the previous study by Suzaku \citep{Nobukawa20}.
All the parameters for MAXI\,J1421$-$613 were fixed to those of the fitting results in the {\tt phabs}$\times${\tt nthComp} model described above. 
The absorption column density $N_{\rm H}$ of the annular emission was fixed to the value of \citet{Nobukawa20}, $N_{\rm H}=3.7\times10^{22}$~cm$^{-2}$, while the parameters of the power law (photon index and normalization) were free. The photon index and unabsorbed flux in the 2--5~keV band of the annular emission were obtained to be $3.9\pm0.2$ and $(10.1\pm0.7)\times10^{-12}$~erg~s$^{-1}$~cm$^{-2}$, respectively. The best-fit results of the annular emission are summarized in table 2. 

\begin{figure}
\centering
\includegraphics[scale=0.12]{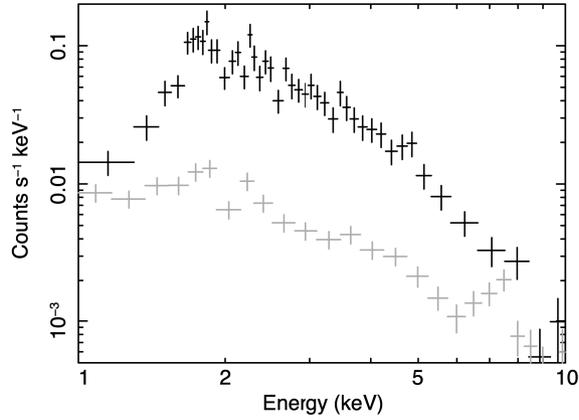}
\caption{Spectrum extracted from the annular region indicated by the two circles in figure 1b (black) and background spectrum extracted from an annular region of radii $6'$--$10'$ centered on MAXI\,J1421$-$613 (gray). The difference in the detector area covered by the two extraction regions was corrected. }
\label{fig:pendulum}
\end{figure}

\begin{table*}
\centering
\caption{Best-fit parameters of the annular emission.}
\begin{tabular}{cc} \hline
Component			& Values   	\\  \hline
$N_{\rm H}$ (cm$^{-2}$)	& $3.7\times10^{22}$ (fixed) \\	
$\Gamma$				& $3.9\pm0.2$		\\
Flux$^{\ast}$	& $(10.1\pm0.7)\times10^{-12}$ \\ \hline
$\chi^2$/d.o.f. 	& $11.04/11$ \\
\hline
\multicolumn{2}{l}{$^{\ast}$   Unabsorbed flux in the 2--5~keV band in the unit of  erg~s$^{-1}$~cm$^{-2}$.} \\
\end{tabular}
\end{table*}

\section{Discussion}
Among the Swift follow-up observations of MAXI J1421$-$613 obtained by PC mode, we analyzed the data not analyzed in \citet{Nobukawa20}. Although the data taken on the same day as the MAXI alert (9 January 2014) were affected by severe pileup, the observation data obtained 9 days after the MAXI alert clearly show the annular emission with a radius of $3'.5$ centered on MAXI\,J1421$-$613 in both the X-ray image and the radial profile. In this section, we compared this result with the previous study of the annular emission observed by Suzaku \citep{Nobukawa20}.

The spectrum of the annular emission observed by Suzaku was represented by the power law of the photon index $\Gamma = 4.2 \pm 0.3$. In this study,  the spectrum of the annular emission extracted from the Swift data was also fitted by the power law of photon index $\Gamma = 3.9 \pm 0.2$.  These $\Gamma$ values are higher than the value of MAXI\,J1421$-$613 itself ($\sim 2.1$; Serino et al., 2015) by $\Delta \Gamma \sim 2$.  In the optically thin limit where  $(1 - {\rm e}^{-\tau})$ can be approximated as $\tau$, the photon energy dependence of the differential cross-section of X-ray dust scattering is given by $E^{-\Gamma}$. Here $\Gamma = 2$ \citep{Draine03}. Thus the annular emission found in this paper would be made by dust scattering. 

The flux of the annular emission observed by Suzaku from 31 January to 3 February 2014 ($(2.5\pm0.1)\times10^{-12}$~erg~cm$^{-2}$~s$^{-1}$; Nobukawa et al., 2020)  is $\sim20$\% of that observed by Swift on 18 January 2014 ($(10.1\pm0.7)\times10^{-12}$~erg~s$^{-1}$~cm$^{-2}$). This would be due to the scattering-angle dependence of the differential cross-section of X-ray dust scattering  \citep{Draine03}. According to equation (5) of \citet{Heinz16}, the differential cross-section is expected to decrease to 15\%--24\% as the radius of the annular emission increases from $3'.5$ (Swift) to $5'.6$ (Suzaku).  The annular emission observed by Swift is considered to be caused by dust scattering of the same outburst as is responsible for the annular emission observed by Suzaku.

In the previous study using Suzaku \citep{Nobukawa20}, the distance to MAXI\,J1421$-$613 was estimated with the two assumptions: (1) the brightest radius $6'$ of the annular emission was made by the outburst at the flux peak on 9 January 2014, and (2) the dust layer responsible for the annular emission is located at the same position as the CO cloud (2.6~kpc or 9~kpc) that is likely to be located in front of MAXI\,J1421$-$613 \citep{Nobukawa20}. The radius of the annular emission ($5'.6$) and the difference in the light traveling times (21--24~days) in the Suzaku observation gave the estimation of the distance to MAXI\,J1421$-$613 of $\sim3$~kpc in the case that the dust layer is located at 2.6~kpc, and $\sim30$~kpc in the case of the dust layer at 9~kpc. The latter is unlikely because MAXI\,J1421$-$613 is beyond the Galaxy. Also, if MAXI\,J1421$-$613 is located at 30 kpc, the flux of the first X-ray burst ($7\times10^{-8}$~erg~s$^{-1}$~cm$^{-2}$) reported by \citet{Serino15} means the luminosity of $3\times10^{39}$~erg~s$^{-1}$, which is an order of magnitude above the Eddington limit for a neutron star. \citet{Nobukawa20} concluded that the distance to MAXI\,J1421$-$613 was  $\sim3$~kpc. 

Table~1 in this paper summarizes Swift's follow-up observations. We found that the small amount of data acquired in PC mode makes it difficult to limit both the distance to MAXI J1421-613 and the dust layer at once. Therefore, in this paper, we estimate the distance to MAXI\,J1421$-$613 with the same assumptions as in the previous study. Based on the equation (8) in \cite{Trumper73}, we used the difference in the light traveling times (9~days), the distance to the dust layer (2.6~kpc or 9~kpc), and the radius of the annular emission ($3'.5$). If the distance to the dust layer is 2.6~kpc,  the calculated distance to MAXI\,J1421$-$613 is $\sim3$~kpc. In the case of the dust layer at 9~kpc, the distance to MAXI\,J1421$-$613 is estimated to be more than 20~kpc. The latter is still unlikely. These results are fully consistent with the previous study using Suzaku. Further observations in the future, when MAXI\,J1421$-$613 exhibits outbursts again, will reveal more details on the distance to the dust layer and that of MAXI\,J1421$-$613. 

\section{Conclusion}
Analyzing the Swift follow-up observation of MAXI\,J1421$-$613 which was conducted by the PC mode on 18 January 2014 (9~days after the MAXI alert), we found the annular emission at the radius of $\sim3.5'$ in the X-ray image and the radial profile. The spectrum of the annular emission was well explained by the absorbed power law with the photon index of $\Gamma = 3.9 \pm 0.2$, which is higher than the value of MAXI\,J1421$-$613 itself ($\sim 2.1$; Serino et al., 2015) by $\Delta \Gamma \sim 2$. These results confirm the annular emission was caused by dust scattering of X-rays from MAXI\,J1421$-$613.  
The flux of the annular emission studied by Suzaku, which observed the source from 31 January to 3 February 2014,  decreased to $\sim20$\% of the one observed by Swift. The flux decrease is explained by the scattering-angle dependence of the differential cross-section of dust scattering. The annular emission observed by Swift would be caused by dust scattering of the same outburst as is responsible for the annular emission observed by Suzaku.
Assuming that the dust layer causing the annular emission found by Swift is located at the same position as the CO cloud in front of  MAXI\,J1421$-$613, the distance to MAXI\,J1421$-$613 was estimated to be $\sim3$~kpc, which is consistent with the value estimated by the Suzaku observation.

\section*{Acknowledgments}
This work was supported by MEXT KAKENHI No.JP20K14491, JP20KK0071 (KN), and JP21K03615 (MN). KKN was also supported by Yamada Science Foundation.

\bibliographystyle{model5-names}
\biboptions{authoryear}
\bibliography{refs}

\end{document}